\documentclass [a4paper]{article} %, fleqn
\newenvironment{affiliations}{\begin{enumerate}}{\end{enumerate}}
\newenvironment{methods}{\section*{Methods}}{}

\usepackage{graphicx} % this is needed for my \documentclass, and doesn't harm nature's

\usepackage{amsmath} %this allows the \eqref command, which nicely puts parentheses around the equation numbers referenced in the text
\usepackage{upgreek}
\usepackage[normal]{subfigure} 
\usepackage{float}
\usepackage{hyphenat}
 
\restylefloat{figure}

\usepackage{layout}%questo permette di variare parametri di layout, come i quattro seguenti
\title{Full stress tensor measurement using colour centres in diamond}

\author{Fabio Grazioso$^1$, Brian R. Patton$^1$, Paul Delaney$^{2,3}$, Matthew L. Markham$^4$,\\ Daniel J. Twitchen$^4$, \& Jason M. Smith$^1$} 

\begin{document}

\maketitle

\begin{affiliations}
\item Department of Materials, University of Oxford, Parks Road, Oxford OX1 3PH, United Kingdom.
\item School of Mathematics and Physics, Queen's University Belfast, Belfast BT7 1NN, Northern Ireland.
\item Tyndall National Institute, University College Cork, Lee Maltings, Prospect Row, Cork, Ireland.
\item Element Six Ltd., King's Ride Park, Ascot, Berkshire SL5 8BP, United Kingdom.
\end{affiliations}

\begin{abstract}
Stress and strain are important factors in determining the mechanical, electronic, and optical properties of materials, relating to each other by the material's elasticity or stiffness. Both are represented by second rank field tensors with, in general, six independent components \cite{Nye}. Measurements of these quantities are usually achieved by measuring a property that depends on the translational symmetry and periodicity of the crystal lattice, such as optical phonon energies using Raman spectroscopy \cite{Loechelt99}, the electronic band gap using cathodoluminescence \cite{Tang94}, photoelasticity via the optical birefringence \cite{Voloshin83}, or Electron Back Scattering Diffraction (EBSD) \cite{Schwartz00}.  A reciprocal relationship therefore exists between the maximum sensitivity of the measurements and the spatial resolution. Furthermore, of these techniques, only EBSD and off-axis Raman spectroscopy allow measurement of all six components of the stress tensor, but neither is able to provide full 3D maps. Here we demonstrate a method for measuring the full stress tensor in diamond, using the spectral and optical polarization properties of the photoluminescence from individual nitrogen vacancy (NV) colour centres. We demonstrate a sensitivity of order 10 MPa, limited by local fluctuations in the stress in the sample, and corresponding to a strain of about 10$^{-5}$, comparable with the best sensitivity provided by other techniques.  By using the colour centres as built-in local sensors, the technique overcomes the reciprocal relationship between spatial resolution and sensitivity and offers the potential for measuring strains as small as 10$^{-9}$ at spatial resolution of order 10 nm. Furthermore it provides a straightforward route to volumetric stress mapping. Aside from its value in understanding strain distributions in diamond, this new approach to stress and strain measurement could be adapted for use in micro or nanoscale sensors.
\end{abstract}
\begin{figure}[!htbp]
\centering
\includegraphics[width=9cm] {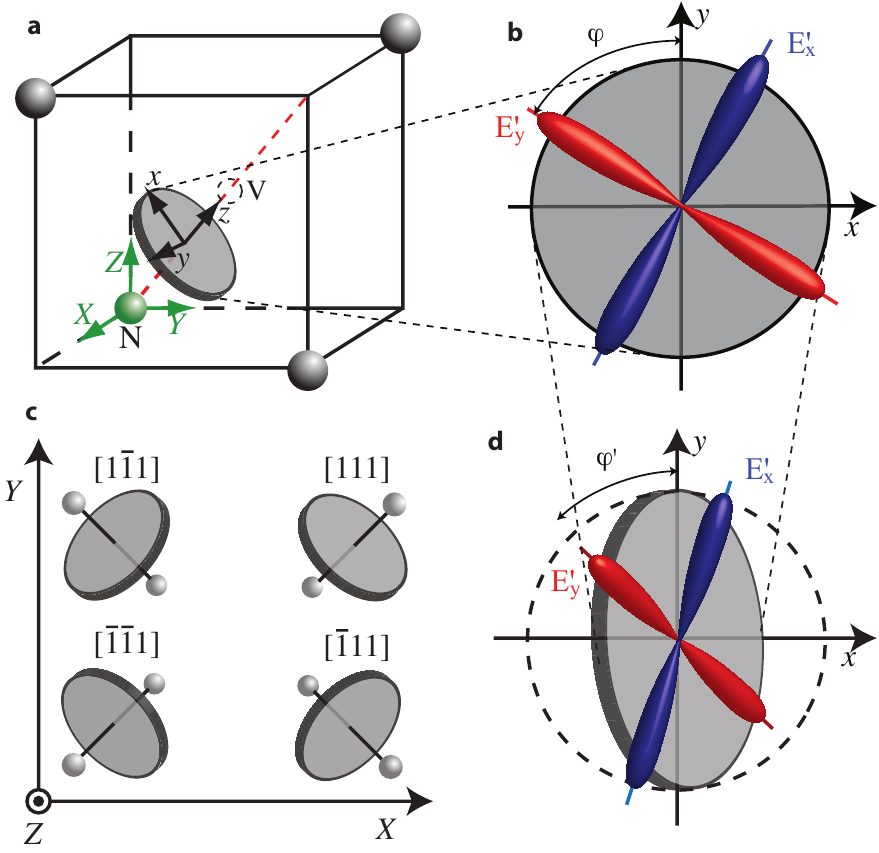}
\caption{ \label{figure1} The NV centre structure: (a) Schematic of the NV structure and definition of NV axes $(x,y,z)$ and crystal axes $(X,Y,Z)$. The cube represents $\frac{1}{8}$ of the cubic unit cell of the diamond crystal. The nitrogen atom is colored green and the vacancy is represented by the dashed circle in the center of the cube. The three carbon atoms adjoining the vacancy are in grey. (b) Rotation of the $E_{x'}$ and $E_{y'}$ orbitals relative to NV axes $x$ and $y$ under a symmetry-breaking  stress. $E_{x'}$ is taken to be the higher energy orbital, consistent with \cite{Tamarat06}. (c) Projection onto the $(001)$ image plane $(XY)$ of the four possible orientations of NVs relative to the crystal axes. (d) Projection of the $E_{x',y'}$ orbitals in (b) onto the image plane, leading to the measured data in figure \ref{PolarPlots}.}
\end{figure}

\vspace{1 cm}
The NV centre is the most common colour centre in diamond, comprising a substitutional nitrogen atom and a vacancy in an adjacent site along one of the four $\left\langle111\right\rangle$ crystal axes, as depicted in figure \ref{figure1}(a). The perturbation of its optical emission by applied uniaxial stress via the piezospectroscopic effect \cite{Kaplyanskii-64a} was characterised in the seminal work of Davies and Hamer in 1976 \cite{Davies76} and used to identify the centre as possessing $C_{3v}$ symmetry and a $E \rightarrow A$ optical transition. In recent years, optical microscopy of individual centres has become popular since they were demonstrated to be exceptionally stable single photon emitters \cite{Kurtsiefer00} and to possess an electronic structure that couples the highly coherent spin state of the centre to its optical transitions, \cite{Jelezko04} thus offering considerable potential for quantum information applications. As a result of this renewed interest, the electronic structure of the colour centre has been studied extensively and is now well understood in terms of the orbital energies, spin-orbit and spin-spin interactions \cite{Manson-06,Batalov09,Doherty11}, and hyperfine coupling. The perturbation of the energy levels by external electric and magnetic fields is also established, leading to demonstrations of the use of NV centres as nanoscale sensors \cite{Balasub08,Dolde11}.

The effect of local stress on a single NV centre can be understood by considering the effect that different `components' of the stress field have on the defect symmetry. To this end we define a set of orthogonal axes $(x,y,z)$ with $z$ parallel to the NV axis and $x$ lying in one of the three mirror planes. For example an NV centre with its ($z$) axis along $[111]$ we choose $x$ to lie along $[\bar{1}\bar{1}2]$ and $y$ along $[1\bar{1}0]$ as shown in figure \ref{figure1}(a). 

The $^3E$ optically excited state of the NV centre has twofold spatial degeneracy and threefold spin degeneracy, and under the $C_{3v}$ symmetry of the defect these six eigenstates are split into four levels by the spin-orbit and spin-spin interactions, with splittings of order a few GHz. Under stress the energies of the $^3E$ states are perturbed. A stress that preserves the $C_{3v}$ symmetry will result in a global shift of the $^3E$ levels, whilst a stress that breaks the symmetry modifies the splittings in the $^3E$ manifold and rotates the orbitals in the $(x,y)$ plane. In this respect the stress perturbation is similar to that of a vector with three components ($\xi_x, \xi_y, \xi_z$) \cite{Tamarat08,Dolde11}. In the limit that the stress perturbation greatly exceeds the spin-orbit and spin-spin splittings the six $^3E$ states form two branches corresponding to orthogonal spatial orbitals $E_{x'}$ and $E_{y'}$, and the perturbation can be expressed as a secular matrix in the $(E_x,E_y)$ basis \cite{Hughes-67}, 

\begin{equation}\label{secular-matrix}
\left(
\begin{array}{cc}
 \alpha + \beta    &   \gamma \\
  \gamma   &  \alpha - \beta
\end{array}
\right)
\end{equation}

where $\alpha$, $\beta$ and $\gamma$ are energy shifts determined by the six components of the stress tensor. The eigenvalues of this secular matrix are the energy changes of the stress-shifted $E$ states, and the eigenvectors provide the orientations of $x'$ and $y'$, the $^3E \rightarrow ^3A$ transition dipoles, in the ($x,y$) plane. The optical transition line therefore becomes a doublet with a mean energy shift of $\alpha$ and a splitting of $\delta = 2\sqrt{\beta^2+\gamma^2}$, and the $E$ dipoles are rotated by an angle $\phi = 1/2$ arctan$(\gamma/\beta)$ relative to $x$ and $y$ (figure \ref{figure1}(b)). 
For a $[111]$ oriented NV centre the energy shifts are given by

\begin{equation} 
\label{equation-15-111}
\begin{array}{l}
\alpha= A_{1}\left(\sigma_{_{XX}} +\sigma_{_{YY}}+ \sigma_{_{ZZ}} \right) + 2A_{2}\left(\sigma_{_{YZ}} + \sigma_{_{ZX}} + \sigma_{_{XY}}   \right) \\
\beta=  B\left( 2\sigma_{_{ZZ}}- \sigma_{_{XX}} - \sigma_{_{YY}} \right) + C\left(2\sigma_{_{XY}} -\sigma_{_{YZ}} - \sigma_{_{ZX}}  \right)\\
\gamma=\sqrt{3} B\left(\sigma_{_{XX}} -\sigma_{_{YY}} \right) + \sqrt{3}C\left(\sigma_{_{YZ}} -\sigma_{_{ZX}} \right) 
\end{array}
\end{equation}

where the parameters $A_1$, $A_2$, $B$, and $C$ have been determined as 1.47, -3.85, -1.04, and -1.69 meV/GPa respectively \cite{Davies76}, and the stress components $\sigma_{ij}$ (where $i,j = X,Y,Z$) are referred to the crystal axes. The equations for an NV centre with one of the three other orientations are obtained by performing simple rotations on the coordinate axes $(X,Y,Z)$ (see supplementary information). 
The equivalent equations to \eqref{equation-15-111} with the stress referred to the NV axes are 

\begin{equation} 
\label{equation4}
\begin{array}{l}
\alpha= A_{1}\left(\sigma_{_{xx}} +\sigma_{_{yy}}+ \sigma_{_{zz}} \right) + A_{2}\left(2\sigma_{_{zz}} - \sigma_{_{xx}} - \sigma_{_{yy}}   \right) \\
\beta=  (B+C)\left( \sigma_{_{xx}}- \sigma_{_{yy}} \right) + \sqrt{2}(2B-C)\sigma_{_{xz}}\\
\gamma=-2(B+C)\sigma_{_{xy}} +\sqrt{2}(2B-C)\sigma_{_{yz}}
\end{array}
\end{equation}

from which it is clear that the global shift $\alpha$ results from any stress that shares the NV symmetry ($\xi_z$), $\beta$ results from stress that shares only the $xz$ mirror plane symmetry ($\xi_x$), and $\gamma$ results from stress with neither the NVs cylindrical nor its $xz$ mirror symmetry ($\xi_y$).

Since the stress tensor contains six independent parameters, measurement of a single NV centre can clearly provide only partial information of the local stress field. To obtain all six components of the tensor it is therefore necessary to measure three NV centres that experience the same stress field but which have  different orientations in the lattice. With $\alpha$, $\beta$ and $\gamma$ determined for three differently oriented colour centres one obtains an overdetermined set of nine linear equations with six unknown variables, from which a unique solution is obtained using the Moore-Penrose pseudoinverse method \cite{Penrose-56}.

\begin{figure}[!htbp]
\centering
\subfigure[\label{PLimage} ]
{\includegraphics[scale=0.8]{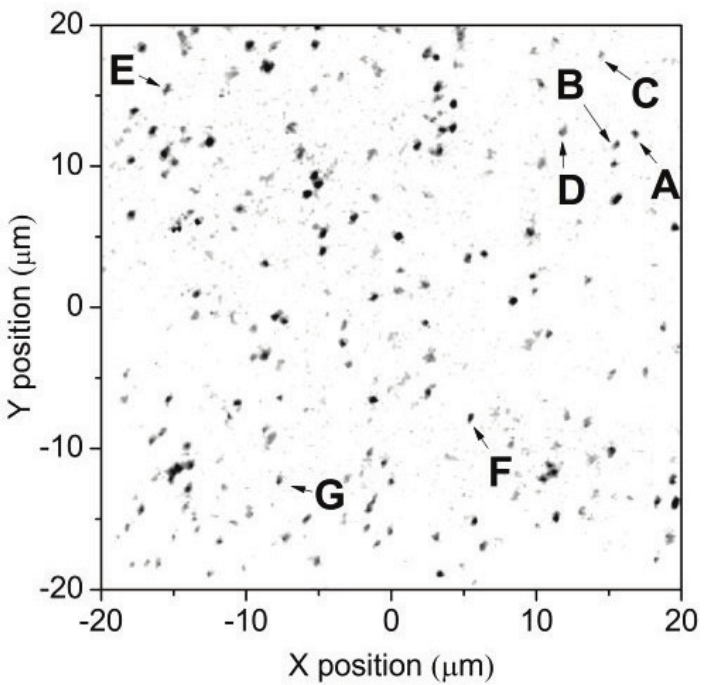}}
%\hspace{5mm}
\subfigure[\label{PLspectrum} ]
{\includegraphics[scale=0.8]{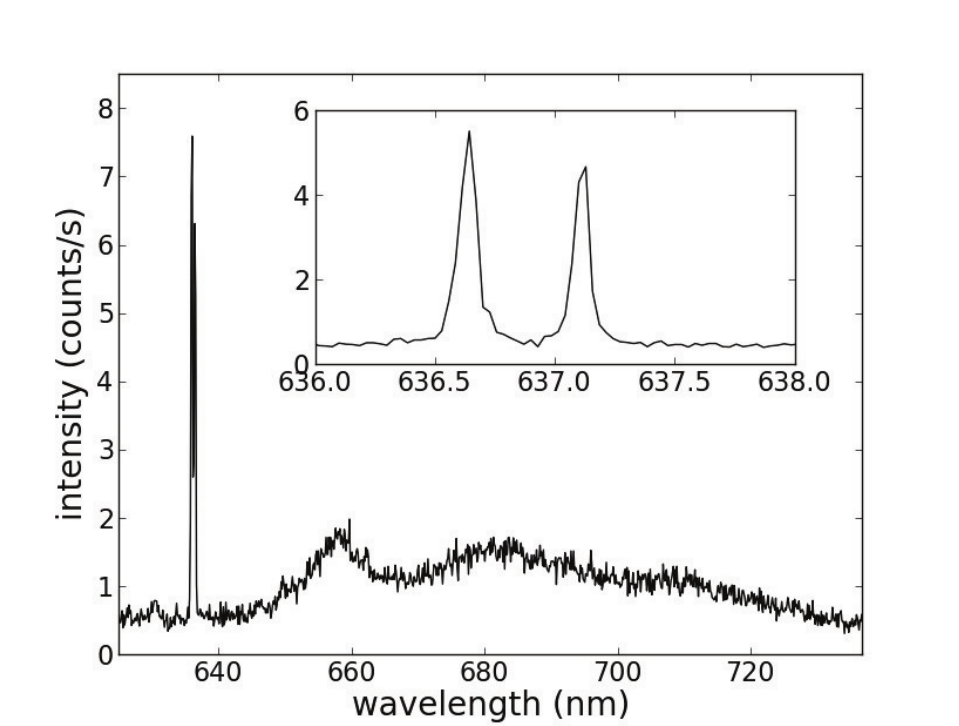}}
\caption{\label{PLdata} Low temperature Photoluminescence from individual nitrogen vacancy centres in highly strained material: (a) PL image of the sample area of interest with the seven NVs used in this study labeled A-G; (b) PL spectrum of a single NV centre with a split zero phonon line (inset).}
\end{figure}

To demonstrate this method we present photoluminescence data from isolated NV centres in a single crystal diamond grown by plasma assisted chemical vapour deposition. A sample was selected which displays isolated NV centres in a region of particularly high grown-in stress (figure \ref{PLimage}), so that at a temperature of 77 K, many of the zero phonon lines reveal clear splittings of up to 0.8 nm or 2.4 meV (figure \ref{PLspectrum}). The viewing axis is $Z\parallel[001]$, such that the integrated emission of a given NV centre is approximately $3:1$ polarized, allowing easy discrimination between NVs with axis parallel to $[111]$ or $[\bar{1}\bar{1}1]$ and those with axis parallel to $[\bar{1}11]$ or $[1\bar{1}1]$ (figure \ref{figure1}(c)). The ambiguity (effectively a mirror in $Z$) in identifying the orientation of a given NV centre is not possible to resolve with this sample geometry, but as we shall see presents only a minor limitation to the results.

The two $E \rightarrow A$ electronic transitions emit linearly polarized photons aligned with the dipole axes. The measured projections of these polarizations onto the $(001)$ image plane allow determination of the angle $\phi$ and thus we obtain $\gamma = \delta \sin(2\phi$) and $\beta = \delta \cos(2\phi)$. The parameter $\alpha$ is taken as the energy shift between the mid point of the two lines and 1.945 eV (637.5 nm).

\begin{figure}[!htb]
\includegraphics[width=9cm]{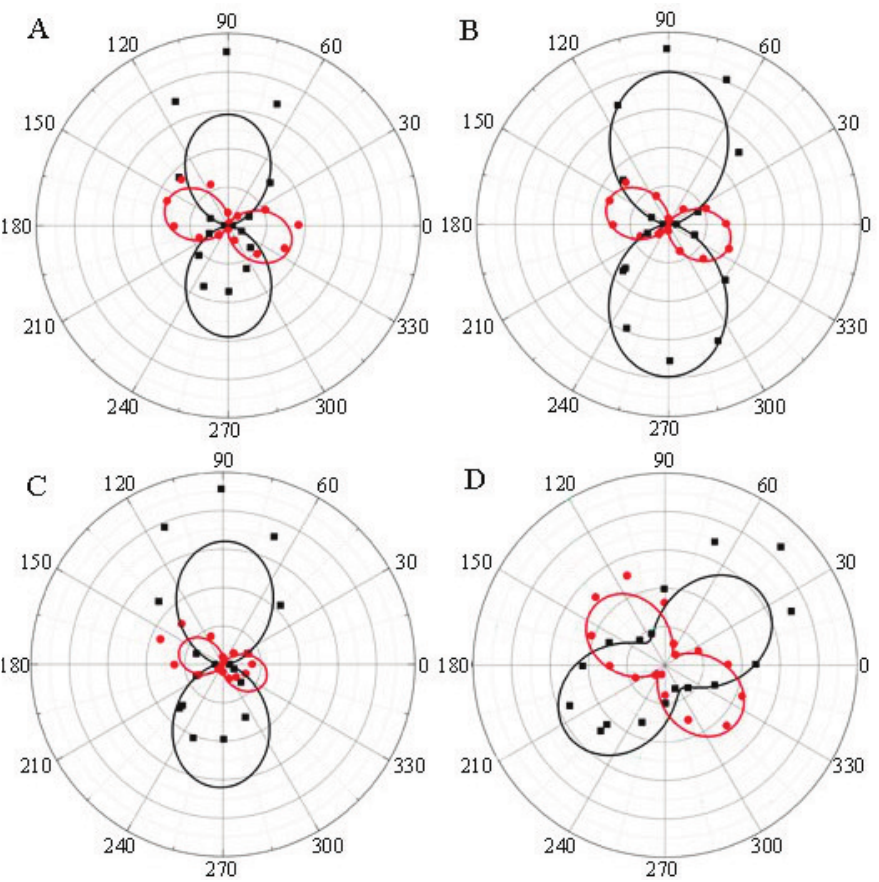}
\centering
\caption{\label{PolarPlots} Measured polarisation data for NVs A, B, C and D. The radius represents the emission intensity and the angle is that of the polarizing filter, $\phi_p$. The solid curves are fits to the function $I_0 + \Delta I cos^2(\phi_p+\phi')$ where $I_0$, $\Delta I$, and $\phi'$ are fitting parameters.}
\end{figure}

As the z-projections of individual NVs are not known we performed measurements on several proximal centres and looked for groupings of parameters to identify different defect orientations. An example of a group of seven NVs studied is shown in figure \ref{PLimage}. Polar plots revealing the optical polarization from each of these centres are shown in figure \ref{PolarPlots}, and the corresponding $\alpha,\beta,\gamma$ values listed in table \ref{table1}.

\begin{table} 
\centering
\begin{tabular}{|l|cccc|}
\hline
defect & angle & $\alpha (meV)$ & $\beta (meV)$ & $\gamma (meV)$ \\
\hline
A & 90 & 1.966 &	1.712	& -0.688\\
B & 90 & 2.359 &	1.391 &	-0.563\\
C & 90 & 2.354 &	1.347	& -0.599\\
D & 0  & 1.322 &	0.495	& 1.376\\
E & 0  & 1.137 &	0.163	& 1.191\\
F & 0  & 1.694 &	0.304	& 1.087\\
G & 0  & 1.740 &	0.354	& 1.052\\
\hline
\end{tabular}
\caption{\label{table1} Polarisation and energy shift data for seven NV centres labeled in figure \ref{PLdata}.}
\end{table}

It can be seen from the tabulated data that in two cases pairs of NVs that are spatially close (B \& C, and F \& G) give very similar values for all parameters measured. The conclusion is that these NVs are co-oriented, and are not affected substantially by local perturbations to the lattice. Of the 22 NVs measured in total for this study, we found six examples of pairs of proximal NVs whose parameters agreed to better than 10$\%$, suggesting that for these NVs the assumption of a dominant long-range stress field is valid. Several NVs also revealed behaviour that was inconsistent with the assumption, presumably resulting from the presence of a strong local stress field due to a nearby defect such as a lattice vacancy, or an electric field due to a trapped charge.

We now use the data for NVs A, (B or C) and D to calculate the local stress tensor in the top right of the image. Choosing the orientations A$\parallel[\bar{1}11]$, (B or C)$\parallel[1\bar{1}1]$ and D$\parallel[111]$ yields the stress tensor
     
\begin{equation} \label{sigma}
\sigma=\left(
\begin{array}{ccc}
	290 (5) & 90 (0) & \pm6 (0)\\
	90 (0) & 758 (0) & \pm45 (0)\\
	\pm6 (0) & \pm45 (0) & -14 (2)\\
\end{array}
\right) \text{MPa}
\end{equation}

where the figures presented are the mean values (standard deviations) for the tensor components calculated using the sets \{A,B,D\} and \{A,C,D\}. The $\pm$ values in  $\sigma_{XZ}$ and $\sigma_{YZ}$ indicate the effect of inverting $Z$ for all three centres. Reversing the choice of orientation for D (only) creates a distinct configuration and results in the tensor

\begin{equation} \label{sigmaprime}
\sigma'=\left(
\begin{array}{ccc}
	259 (1) & 98 (2) & \pm44 (2)\\
	98 (2) & 758 (4) & \pm27 (2)\\
	\pm44 (2) & \pm27 (2) & -31 (6)\\
\end{array}
\right) \text{MPa}
\end{equation}

The most significant differences between $\sigma$ and $\sigma'$ are in the tensor components containing $Z$, corresponding to the $Z$ reflection of centre D. We note however that the choice of NV B or C makes very little difference to the results in either case, demonstrating that the difference in the stress experienced between these two NV's is of the order of a few MPa at most. That the standard deviations are consistently lower in $\sigma$ than in $\sigma'$ may further suggest that this choice of orientation for D is the correct one of the two.

From a calculated stress tensor we can use equations \eqref{equation-15-111} to recalculate $\alpha, \beta$ and $\gamma$ for each of the NV orientations thus providing a convenient test of the consistency of the measurement. Tensor $\sigma$ yields the results presented in table \ref{table2}. 

\begin{table}[!htbp]
\centering
\begin{tabular}{|l|cccc|}
\hline
NV axis & angle & $\alpha (meV)$ & $\beta (meV)$ & $\gamma (meV)$ \\
\hline
$[111]$                     & 0  & 1.219 &	0.729	& 0.957\\
$[1\bar{1}1]$               & 90 & 2.511 &	1.358	& -0.993\\
$[\bar{1}\bar{1}1]$        & 0  & 0.434 &	0.901	& 0.731\\
$[\bar{1}11]$               & 90 & 1.914 &	1.489 &	-0.695\\
      
\hline
\end{tabular}
\caption{\label{table2} Recalculated energy perturbations for the four NV orientations using stress tensor $\sigma$.}
\end{table}

From these results we can clearly identify the $[111]$ values as relating to NV D, $[1\bar{1}1]$ relating to NV's B and C, and $[\bar{1}11]$ relating to NV A.  The discrepancies between the recalculated values and measured values are of order 10$\%$, consistent with the observed variation in parameters described previously. Noteworthy is the absence of any NVs with parameters comparable to the predictions for the $[\bar{1}\bar{1}1]$ orientation. We attribute the orientations of all three (more distant) NV's E, F and G as $[111]$ the differences between their parameters and those of NV D being due to longer range changes in the stress field.

\section*{Discussion}

The tensors in equations \eqref{sigma} and \eqref{sigmaprime} indicate a stress that is primarily biaxial in the $(001)$ plane, with principal axes closely aligned to the $[100]$ and $[010]$ crystal axes. That there is very little tensile or compressive stress along the [001] axis is not surprising as this is the growth axis of the crystal and the NV defects measured here are $\sim 50\ \upmu$m below the surface. The apparent absence of any NVs oriented parallel to $[\bar{1}11]$ may be a result of both the stressed lattice and the NV defects having been generated during the CVD grown process, whereupon there are likely to be preferred and inhibited directions for defect orientation \cite{Newton}.   

We now discuss issues related to the sensitivity of the technique. For simplicity we have presented only the theory applicable to a stress perturbation larger than 10 MPa, equivalent to the $\sim$10 GHz splitting of the unperturbed $^3E$ level, and have treated the spin levels as degenerate. Smaller stresses can equally be determined by measuring in more detail the spectrum of the $^3E$ state and using the full theoretical description that includes the spin-orbit and spin-spin interactions \cite{Batalov09,Doherty11}. The sensitivity is also clearly limited by the spectral resolution of the measurements, which is ultimately determined by the transition line widths. These are a strong function of temperature, so that the sensitivity increases as the sample is cooled. Line widths approaching 10 MHz have been observed using photoluminescence excitation spectroscopy at 2.8 K \cite{Tamarat06}, corresponding to stress magnitudes of order kPa, or strains of order 10$^{-9}$. Although at such extreme sensitivity local variations in the built-in lattice stress would certainly dominate, \textit{changes} in strain of this magnitude may ultimately be detectable. At higher temperatures the line widths broaden rapidly due to the dynamic Jahn Teller effect \cite{Fu09}, and at 77 K they are limited to about 10 GHz corresponding to stresses of order 10 MPa. At 300 K the zero phonon line is generally not clearly visible, and so stress mapping under ambient conditions would require utilisation of the stress perturbation to the electron spin state of the centre, measured using optically detected magnetic resonance (ODMR). The spin-orbit interaction introduces a stress component to the spin splittings in both the orbital doublet excited state \cite{Fuchs08} and, through inter level mixing, the orbital singlet ground state \cite{Dolde11}. Although the stress dependence of the spin transitions is much stronger in the excited state, the ODMR line width is limited by the optical relaxation time ($\simeq$12 ns) thereby limiting the sensitivity of excited state techniques. Ground state ODMR probing the spin transition at 2.88 GHz is therefore likely to be the more sensitive method. In this way, Dolde et al have demonstrated the use of single NV centres as electric field sensors with a dc sensitivity of 891 $\pm20$ V cm$^{-1}$ Hz$^{-\frac{1}{2}}$, equivalent to a stress of about 90 kPa Hz$^{-\frac{1}{2}}$. By combining data from three differently oriented NV centes the full stress tensor can similarly be extracted (see supplementary info for details).

Volumetric imaging of the stress tensor is possible by recording a 3-D distribution of defects. With a standard scanning confocal optical microscope this can provide sub-micrometre spatial resolution per defect, therefore up to $\sim1$ $\upmu$m resolution for the full tensor. However NV centres in diamond are also ideally suited to super-resolution optical techniques such as stimulated emission depletion microscopy \cite{Rittweger09} that utilise an intense depopulating beam and require highly robust photon emitters. Such techniques are entirely compatible with the methodology described here and could yield spatial resolution on a length scale of a few nanometres for systematic investigations of local stress fields. We note that such spatial resolution would be fundamentally incompatible with the various existing approaches to stress/strain measurement. EBSD, Raman, and band gap luminescence all interrogate quantities that depend on the periodicity of the crystal lattice and can therefore only measure strain averaged over a volume, or length, that is large compared with the lattice spacing. The use of point defects as sensors therefore opens up new possibilities for stress/strain measurement such as measuring lattice scale fluctuations in stress caused by nearby defects or surfaces, or for use in nanoscale sensors, for example in nano-electromechanical systems (NEMS).    

\begin{methods}

The sample studied was a high purity, [001] oriented type IIa diamond grown by microwave plasma-assisted chemical vapour deposition (CVD) \cite{Balmer09}. Homoepitaxial CVD diamond growth was performed using a selected high pressure high temperature grown substrate, which had been carefully processed prior to growth in order to minimize the dislocation density in the grown CVD diamond. The resultant diamond had a nitrogen concentration of $<$ 5 ppb as measured by ensemble electron paramagnetic resonance (EPR) and a boron concentration of $<$ 0.5 ppb as measured by secondary ion mass spectrometry (SIMS). After growth the sample was mechanically polished to have a surface RMS roughness $<$ 1 nm over a 1 $\times$ 1 $\upmu$m area.

Low temperature photoluminescence microscopy and spectroscopy was recorded using a custom-made fiber-coupled scanning confocal microscope \cite{Grazioso-10}. Photon counting for image generation utilised a Perkin Elmer SPCM. Spectra were recorded using an Acton SP 500 monochromator and Princeton Spec 10:100B CCD camera. Polarization dependence of emission was measured by placing a linear optical polarizer in front of the collection aperture of the microscope. A slight wedge in the plate leads to partial misalignment of the microscope as the polarizer is rotated, resulting in asymmetry in some of the data in figure \ref{PolarPlots}. 

\end{methods}

\appendix

\section{Supplementary information}

\subsection{ $\alpha$, $\beta$ and $\gamma$ for all NV orientations}

The expressions for $\alpha$, $\beta$ and $\gamma$ depend on the orientation of the NV centre. Since the NV axis can take one of the four $<111>$ directions (the order of the nitrogen and vacancy sites is not important) we require four sets of equations. To calculate these we start with the equations for the $[111]$ oriented defect with the stress tensor components referred to the crystal axes $X$, $Y$, and $Z$ (see equation \eqref{equation-15-111}) and apply successive rotations of $\frac{\pi}{2}$ around the $z$ axis to the coefficients matrix $M=\{\sigma_{ij}\}$. Since the stress component $\sigma_{ij}$ transforms under symmetry operations as the product $ij$, the result of these rotations can be seen by the successive mappings $X\mapsto Y$ and $Y \mapsto -X$. The full set of equations is presented below.

For the $[111]$ oriented NV centre:
\begin{equation} \label{piezosp-equations}
\begin{array}{l}
\alpha= A_{1}\left(\sigma_{_{XX}} +\sigma_{_{YY}}+ \sigma_{_{ZZ}} \right) + 2A_{2}\left(\sigma_{_{YZ}} + \sigma_{_{ZX}} + \sigma_{_{XY}}   \right) \\
\beta=  B\left( 2\sigma_{_{ZZ}}- \sigma_{_{XX}} - \sigma_{_{YY}} \right) + C\left(2\sigma_{_{XY}} -\sigma_{_{YZ}} - \sigma_{_{ZX}}  \right)\\
\gamma=\sqrt{3} B\left(\sigma_{_{XX}} -\sigma_{_{YY}} \right) + \sqrt{3}C\left(\sigma_{_{YZ}} -\sigma_{_{ZX}} \right) .
\end{array}
\end{equation}

For a $[\bar{1}11]$ oriented NV centre:
\begin{equation} 
\begin{array}{l}
\alpha= A_{1}\left(\sigma_{_{XX}} +\sigma_{_{YY}}+ \sigma_{_{ZZ}} \right) + 2A_{2}\left(- \sigma_{_{YZ}} + \sigma_{_{ZX}} - \sigma_{_{XY}}   \right) \\
\beta=  B\left( 2\sigma_{_{ZZ}}- \sigma_{_{XX}} - \sigma_{_{YY}} \right) + C\left(-2\sigma_{_{XY}} +\sigma_{_{YZ}} - \sigma_{_{ZX}}  \right)\\
\gamma=\sqrt{3} B\left(-\sigma_{_{XX}} +\sigma_{_{YY}} \right) + \sqrt{3}C\left(\sigma_{_{YZ}} +\sigma_{_{ZX}} \right) .
\end{array}
\end{equation}

For the $[\bar{1}\bar{1}1]$ oriented NV centre:
\begin{equation} 
\begin{array}{l}
\alpha= A_{1}\left(\sigma_{_{XX}} +\sigma_{_{YY}}+ \sigma_{_{ZZ}} \right) + 2A_{2}\left(-\sigma_{_{YZ}} - \sigma_{_{ZX}} + \sigma_{_{XY}}   \right) \\
\beta=  B\left( 2\sigma_{_{ZZ}}- \sigma_{_{XX}} - \sigma_{_{YY}} \right) + C\left(2\sigma_{_{XY}} +\sigma_{_{YZ}} + \sigma_{_{ZX}}  \right)\\
\gamma=\sqrt{3} B\left(\sigma_{_{XX}} -\sigma_{_{YY}} \right) + \sqrt{3}C\left(-\sigma_{_{YZ}} +\sigma_{_{ZX}} \right) .
\end{array}
\end{equation}

For a $[1\bar{1}1]$ oriented NV centre:
\begin{equation} 
\begin{array}{l}
\alpha= A_{1}\left(\sigma_{_{XX}} +\sigma_{_{YY}}+ \sigma_{_{ZZ}} \right) + 2A_{2}\left(+ \sigma_{_{YZ}} - \sigma_{_{ZX}} - \sigma_{_{XY}}   \right) \\
\beta=  B\left( 2\sigma_{_{ZZ}}- \sigma_{_{XX}} - \sigma_{_{YY}} \right) + C\left(-2\sigma_{_{XY}} -\sigma_{_{YZ}} + \sigma_{_{ZX}}  \right)\\
\gamma=\sqrt{3} B\left(-\sigma_{_{XX}} +\sigma_{_{YY}} \right) + \sqrt{3}C\left(-\sigma_{_{YZ}} -\sigma_{_{ZX}} \right) .
\end{array}
\end{equation}

\subsection{The effect of different viewing orientations}
\begin{figure}[!htbp]
\centering
\subfigure[{view along $[111]$ direction}]
{\includegraphics[scale=1]{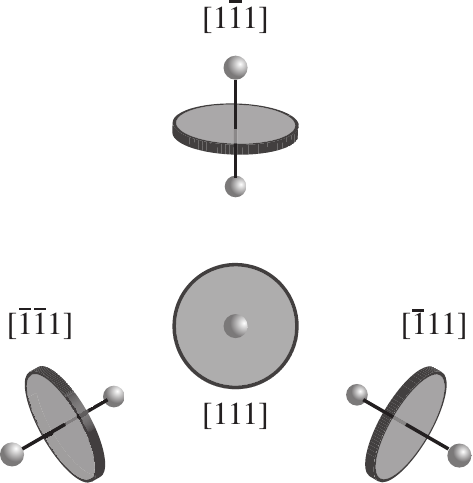} \label{subfigure1}  }
\hspace{10mm}
\subfigure[{view along [110] direction}]
{\includegraphics[scale=1]{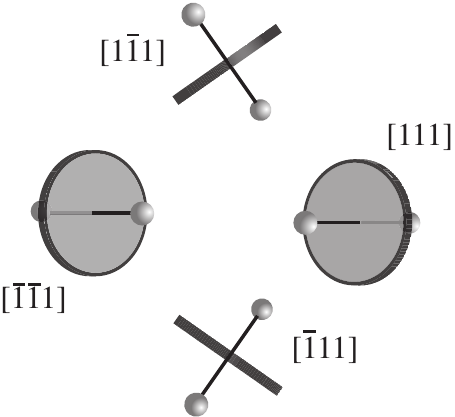} \label{subfigure2}  }
\caption{\label{figure-label} {Projections of the orbital planes of the four different NV orientations when viewed along (a) $[111]$ and (b) $[111]$ direction respectively.}}
\end{figure}

The use of a $[001]$ oriented sample in this study derives from the standard growth conditions of high purity CVD diamond and the ready availability of a suitable sample for study. However as mentioned earlier this viewing axis results in an unavoidable ambiguity in the orientation of any given NV centre, effectively a mirror in $Z$, so that for a given NV centre we can not distinguish between $[111]$ and $[\bar{1}\bar{1}1]$, or between $[\bar{1}11]$ and $[1\bar{1}1]$ orientations. As discussed in the text, we can in some cases make good estimates of which NVs share the same orientation based on their common behaviour in a uniform stress field, but it is not possible to identify unambiguously what that orientation is.
There are two ways of avoiding this problem and thereby providing unambiguous identification of the NV orientations and thus the full stress tensor, subject only to the uniformity of the stress field. The first is to choose a viewing axis that provides straightforward identification of the NV orientation. For example, a $[111]$ viewing axis results in the projections of the orbital planes onto the focal plane as shown in figure \ref{subfigure1}, with four different `ellipses' for the four different NV orientations. The result is that the dependence of PL intensity on excitation polarisation allows simple determination of the orientation of each NV centre.

Conversely, other viewing axes are less conducive to the stress tensor evaluation. As shown in figure \ref{subfigure2}, a $[110]$ orientation results in a twofold ambiguity in orientation as does $[001]$. But, more seriously, it does not allow for measurement of the angle of rotation of the $E_{x'},E_{y'}$ eigenstates away from the $x,y$ axes for two of the orientations since the orbital plane is viewed edge-on. A $[110]$ viewing axis can not therefore be used to determine the full stress tensor by this method.

The second approach is to resolve the ambiguity with a $[001]$ viewing axis, by applying a field which breaks the mirror symmetry in $Z$ (or in both $[110]$ and $[\bar{1}10]$). This could be achieved by the temporary application of an electric field, under which the different orientations would be expected to respond differently. For instance an electric field aligned along a $\left\langle111\right\rangle$ axis would allow identification of NV centres aligned parallel with that axis, as they would display a global shift of the optical transition energies but no change in splitting or polarization.

\subsection{Stress tensor measurement using spin resonance}
The $A_2$ electronic ground state of the NV centre has a non-zero permanent dipole as a result of a small amount of mixing with the $E$ symmetry excited states, which gives rise to a stress and electric field dependence of the spin transitions. The components of this dipole aligned parallel and perpendicular to the NV axis have been measured as $d_{gs}^{\parallel}/h= 0.35\pm0.02 \textnormal{\,Hz cm V}^{-1}$ and $d_{gs}^{\perp}/h= 17\pm3 \textnormal{\,Hz cm V}^{-1}$ \cite{VanOort90}.
The Hamiltonian for the ground state in the presence of local stress, plus electric and magnetic fields \cite{Mims72, VanOort90, Dolde11} is given by

\begin{equation} \label{equation5}
\begin{split}
H_{gs}&=(hD_{gs}+d_{gs}^{\parallel}\Pi_z)[S_z^2-\frac{1}{3}S(S+1)]+\mu_{_B}g_e \mathbf{S} \times \mathbf{B}\\
& \quad -d_{gs}^{\perp}[\Pi_x(S_x^2-S_y^2)+\Pi_y(S_xS_y+S_yS_x)]
\end{split}
\end{equation}

where as defined in figure \ref{figure1}, $z$ is along the NV axis and $x$ is in one of the three mirror planes. $D_{gs}$ is the spin-spin interaction strength, $\mu_{_B}$ is the Bohr magneton, $g_e$ is the electron g-factor. $\Pi_x, \Pi_y$, and $\Pi_z$ are the components of a vector that sums the electric field and the effect of stress on a single NV centre, and $S_x$, $S_y$ and $S_z$ are the triplet spin operators, shown below in the $(x,y,z)$ basis

\begin{equation} \label{equation6}
\begin{array}{lcr}
S_x=\hbar\left(\begin{array}{ccc}
0 & 0 & 0\\
0 & 0 & 1\\
0 & 1 & 0
\end{array}\right)
& 
S_y=\hbar\left(\begin{array}{ccc}
0 & 0 & -i\\
0 & 0 & 0\\
i & 0 & 0
\end{array}\right)
&
S_z=\hbar\left(\begin{array}{ccc}
0 & 1 & 0\\
1 & 0 & 0\\
0 & 0 & 0
\end{array}\right)
\end{array}
\end{equation}

which result in spin eigenstates 
\begin{equation} \label{equation7}
\begin{array}{lcr}
\left|x\right\rangle=\left(\begin{array}{c}
1\\0\\0
\end{array}
\right)
&
\left|y\right\rangle=\left(\begin{array}{c}
0\\i\\0
\end{array}
\right)
&
\left|z\right\rangle=\left(\begin{array}{c}
0\\0\\1
\end{array}
\right)
\end{array}
\end{equation}
while $S(S+1)=S_x^2+S_y^2+S_z^2=2$. Matrix elements can thus be calculated using equations \eqref{equation5}, \eqref{equation6}, and \eqref{equation7} to write the Hamiltonian in the $(x,y,z)$ basis 
\begin{equation}
H_{gs}=\left(
\begin{array}[pos]{ccc}
	hD_{gs}+d_{gs}^{\parallel}\Pi_z+d_{gs}^{\perp}\Pi_x & -d_{gs}^{\perp}\Pi_y & 0\\
	-d_{gs}^{\perp}\Pi_y & hD_{gs}+d_{gs}^{\parallel}\Pi_z-d_{gs}^{\perp}\Pi_x & 0\\
	0 & 0 & -2(hD_{gs}+d_{gs}^{\parallel}\Pi_z)
\end{array}
\right)
\end{equation}
where we have taken the case of zero magnetic field for simplicity (we note that, as pointed out by Dolde et al in \cite{Dolde11}, care must be taken to minimise the magnetic field, especially its $z$ component, to achieve high sensitivity in these measurements). The effect of the combined stress and electric field on the transitions between the $z$ spin state and the other two eigenstates is therefore described by the secular matrix in the spin $(x,y)$ basis
\begin{equation} \label{secular-matrix-appendix}
\left(
\begin{array}{cc}
 \alpha_{gs} + \beta_{gs}    &   \gamma_{gs} \\
  \gamma_{gs}   &  \alpha_{gs} - \beta_{gs}
\end{array}
\right)
\end{equation}
where
\begin{equation}
\begin{array}{l}
\alpha_{gs}=3d_{gs}^{\parallel}\Pi_z\\
\beta_{gs}=d_{gs}^{\perp}\Pi_x\\
\gamma_{gs}=-d_{gs}^{\perp}\Pi_y
\end{array}
\end{equation}

Equation \eqref{secular-matrix-appendix} above can be viewed in direct analogy with the secular matrix for the excited state as described in equation \eqref{secular-matrix}.
In the absence of an electric field the same symmetry arguments that lead to equations \eqref{equation-15-111} and \eqref{equation4}, expressing $\alpha, \beta$ and $\gamma$ for the electric dipole transitions in terms of the stress tensor components and coefficients $A_1, A_2, B \textnormal{ and } C$, also apply here. $\alpha_{gs}, \beta_{gs}$ and $\gamma_{gs}$ are therefore related in the same way to the six independent tensor components via four equivalent constants $A_{1\_gs}$, $A_{2\_gs}$, $B_{gs}$, and $C_{gs}$. This allows us to separate the stress contribution from the electric field contribution and write 

\begin{equation} 
\begin{array}{l}
\alpha_{gs}= A_{1\_gs}\left(\sigma_{_{xx}} +\sigma_{_{yy}}+ \sigma_{_{zz}} \right) + A_{2\_gs}\left(2\sigma_{_{zz}} - \sigma_{_{xx}} - \sigma_{_{yy}}\right) + 3d_{gs}^{\parallel}E_z \\
\beta_{gs}=  (B_{gs}+C_{gs})\left( \sigma_{_{xx}}- \sigma_{_{yy}} \right) + \sqrt{2}(2B_{gs}-C_{gs})\sigma_{_{xz}}+d_{gs}^{\perp}E_x\\
\gamma_{gs}=-2(B_{gs}+C_{gs})\sigma_{_{xy}} +\sqrt{2}(2B_{gs}-C_{gs})\sigma_{_{yz}}-d_{gs}^{\perp}E_y.
\end{array}
\end{equation}
For the $[111]$ oriented centre, with stress and electric field now referred to the crystal axes, we have a set of three equations analogous to those of equation (2),
\begin{equation} 
\begin{array}{l}
\alpha_{gs}= A_{1\_gs}\left(\sigma_{_{XX}} +\sigma_{_{YY}}+ \sigma_{_{ZZ}} \right) + 2A_{2\_gs}\left(\sigma_{_{YZ}} + \sigma_{_{ZX}} + \sigma_{_{XY}}   \right)+\sqrt{3}d_{gs}^{\parallel}\left(E_{_X}+E_{_Y}+E_{_Z}\right) \\
\beta_{gs}=  B_{gs}\left( 2\sigma_{_{ZZ}}- \sigma_{_{XX}} - \sigma_{_{YY}} \right) + C_{gs}\left(2\sigma_{_{XY}} -\sigma_{_{YZ}} - \sigma_{_{ZX}}  \right)+ \frac{d_{gs}^{\perp}}{\sqrt{6}}\left(-E_{_X}-E_{_Y}+2E_{_Z}\right)\\
\gamma_{gs}=\sqrt{3} B_{gs}\left(\sigma_{_{XX}} -\sigma_{_{YY}} \right) + \sqrt{3}C_{gs}\left(\sigma_{_{YZ}} -\sigma_{_{ZX}} \right)-\frac{d_{gs}^{\perp}}{\sqrt{2}}\left(E_{_X}-E_{_Y}\right) .
\end{array}
\end{equation}
with the permutations for the other NV orientations as in equation \eqref{piezosp-equations}. For a single NV centre, $\alpha_{gs}, \beta_{gs}$ and $\gamma_{gs}$ can be measured from optically detected magnetic resonance data by the shift and splitting in the spin transitions plus the orientation of the transition-inducing microwave field. The same pseudoinverse method can thus be applied to determine the full stress tensor from room temperature spin resonance data, once the constants $A_{1\_gs}$, $A_{2\_gs}$, $B_{gs}$, and $C_{gs}$ have been established.

\paragraph{Acknowledgements} The authors would like to thank A. J. Wilkinson for helpful discussions regarding stress and strain measurement techniques. The work was funded by the United Kingdom Engineering and Physical Sciences Research Council through the Quantum Information Processing Interdisciplinary Research Collaboration (grant reference GR/S822176/01). FG and JMS acknowledge additional support from Hewlett Packard Ltd.

\paragraph{Competing Interests} The authors declare that they have no competing financial interests.

 \paragraph{Correspondence} Correspondence and requests for materials should be addressed to\\ J. M. Smith.~(email: \textnhtt{jason.smith@materials.ox.ac.uk}).

\end{document}